# Machine learning reconstruction of digit bone Raman spectra enables noninvasive transcutaneous detection of systemic osteoporosis

MOHAMMAD HOSSEINI,[1,5] SADIA AFRIN,[2] ANTHONY YOSICK,[2,3]

HANI AWAD,[2,3,4] AND ANDREW J. BERGER[1,2,6]

[1]The Institute of Optics, University of Rochester, 275 Hutchison Rd, Rochester, NY 14620, USA

[2]Department of Biomedical Engineering, University of Rochester, 207 Robert B. Goergen Hall, Rochester, NY 14620, USA

[3]The Center for Musculoskeletal Research, University of Rochester Medical Center, 601 Elmwood Avenue, Box 665, Rochester, NY 14642, USA

[4]Department of Orthopaedic, University of Rochester Medical Center, 601 Elmwood Avenue, Box 665, Rochester, NY 14642, USA

[5]shossei3@ur.rochester.edu

[6]andrew.berger@rochester.edu

**Abstract:** Osteoporosis, a major global epidemic, often goes undetected until a fracture occurs, largely due to poor access to screening using gold standard methods, such as dual-energy X-ray absorptiometry (DXA). As a potential nonionizing radiation alternative, we present a transcutaneous spatially offset Raman spectroscopy (SORS) approach combined with machine learning (ML) to recover bone spectra through overlying soft tissue and extract diagnostic information. In a human cadaveric study spanning normal, osteopenic, and osteoporotic donors, we acquired paired Raman measurements from transcutaneous fingers at multiple spatial offsets (0, 3, and 6 mm) and from the corresponding exposed finger bones. Using this paired dataset, supervised machine-learning models were trained to reconstruct exposed-bone Raman spectra from transcutaneous measurements, enabling direct recovery of bone biochemical signatures from transcutaneous tissue. The ML–predicted bone spectra preserved physiologically meaningful Raman features and demonstrated statistically significant differences between normal and osteoporotic groups across four key Raman-derived metrics ($p \leq 0.05$), representing, to our knowledge, the first demonstration of transcutaneous Raman discrimination between clinically established bone-health categories in a human cadaveric study. The ML-predicted spectra further correlated with distal-radius DXA T-scores ($r = 0.73$, $\text{RMSE}_{CV} \approx 1.4$), approaching the performance achieved using exposed-bone measurements ($r = 0.9$, $\text{RMSE}_{CV} \approx 0.8$). Finally, preliminary *in vivo* measurements from two volunteers revealed clear bone-related transcutaneous spectral features consistent with cadaveric data, supporting translational feasibility. Together, these results establish a foundation for nonionizing radiation, transcutaneous Raman assessment of bone health using supervised spectral extraction from accessible measurement sites.

## 1    Introduction

Osteoporosis is a major global health epidemic, affecting hundreds of millions of individuals and substantially increasing the risk of fractures, morbidity, and mortality [1]. Osteoporosis and osteopenia are characterized by low bone mass and micro-architectural deterioration of bone tissue [2], which is clinically quantified using the T-score, defined as the number of standard deviations by which an individual's bone mineral density (BMD) differs from that of a healthy young adult of the same sex. The World Health Organization (WHO) has established classifications of bone loss severity as normal (N, T-score > -1), osteopenia (OPE, -2.5 < T-score $\leq$ -1) and osteoporosis (OP, T-score $\leq$ -2.5) [3]. Dual-energy X-ray absorptiometry (DXA), the current clinical gold standard, measures only bone mineral density and does not capture bone quality, a broader concept encompassing bone architecture, material properties, and remodeling dynamics [4]. As a result, DXA explains only 60–70% of fracture risk [5].

Furthermore, changes in DXA-measured BMD often underpredict the antifracture efficacy of osteoporosis therapies, many of which reduce fracture risk primarily by improving bone quality rather than density alone [6]. DXA also has notable practical limitations, including reliance on ionizing radiation, limited accessibility, and low screening compliance [7], [8]. Consequently, many patients remain undiagnosed and untreated until a fracture occurs, resulting in poor outcomes and rising healthcare costs. These limitations underscore the urgent need for safe, accessible, and noninvasive tools capable of assessing bone quality beyond density alone.

Raman spectroscopy has emerged as a powerful technique for probing the molecular composition and structural organization of materials, including biological tissues [9]. By measuring inelastic scattering of monochromatic light, Raman spectroscopy provides a biochemical fingerprint sensitive to both mineral and organic matrix constituents [10]. In exposed bone, Raman-derived markers such as mineral-to-matrix ratios, carbonate substitution, collagen crosslinking, and crystallinity have been strongly associated with mechanical competence, fracture toughness, aging, disease progression, BMD, and treatment response [11], [12], [13], [14], [15], [16], [17], [18]. Direct transcutaneous Raman assessment of bone, however, remains technically challenging due to overwhelming spectral contributions from overlying tissues such as skin, adipose, and muscle, as well as interference from fluorescence and scattering losses. Early feasibility studies demonstrated that bone Raman signals could be detected through murine skin using approaches such as time-resolved transcutaneous Raman spectroscopy and optical clearing, enhanced transcutaneous Raman measurements [19], [20].

These limitations have been largely overcome by the introduction of spatially offset Raman spectroscopy (SORS), a deep Raman technique that suppresses surface-layer signals and selectively enhances Raman contributions from subsurface layers in diffusely scattering media by spatially separating the illumination and collection zones on the sample surface [21], [22]. This approach has proven highly effective for noninvasive transcutaneous assessment of bone composition and quality in both animal studies [23], [24] and human subjects [25], [26]. Several studies have investigated the spatial offsets required for optimal subsurface signal recovery across different applications [27], [28], [29], [30]. In particular, prior work from our group demonstrated that spatial offsets of approximately 3–6 mm provide the strongest bone-specific features and highest signal-to-noise ratio (SNR) in measurements of the human hand, thereby maximizing sensitivity to the underlying mineralized tissue [31]. More recently, our group demonstrated using exposed-bone measurements that a 3-mm SORS offset yields significantly improved biochemical discrimination compared with conventional surface (0-mm) Raman measurements, further underscoring the importance of appropriate spatial offset selection for bone-specific analysis[32].

A remaining major challenge in transcutaneous SORS is the reliable extraction of weak bone-specific Raman signatures from spectra dominated by overlying tissues, fluorescence background, and photon scatter. Although SORS enhances subsurface sensitivity, the measured spectra still contain complex mixtures of skin, adipose, and bone signals, necessitating robust spectral decomposition strategies. Only a limited number of studies have addressed this problem through advances in probe design, data processing, and multivariate spectral unmixing algorithms. Early work applied band-target entropy minimization (BTEM) and related multivariate curve resolution approaches to isolate bone features from *in vivo* murine SORS measurements [33]. Subsequent method from our group introduced simultaneous, over constrained, library-based decomposition (SOLD) to improve separation of bone and soft-tissue signals and to detect age- and disease-related spectral differences in transcutaneous murine measurements [34]. Additional contributions demonstrated the utility of multivariate decomposition techniques, including BTEM, parallel factor analysis (PARAFAC), and related algorithms, for resolving overlapping subsurface components in human SORS data [35]. More recently, adaptive BTEM approaches have been developed to further optimize bone-signal recovery by dynamically refining band targets based on the spectral characteristics of each measurement [36]. Together, these studies highlight both the promise and the current limitations of spectral unmixing for transcutaneous bone SORS and underscore the need for improved strategies for bone-signal isolation.

Alongside these spectral unmixing approaches, machine learning (ML) techniques have increasingly been applied to Raman spectra of bone to predict biomechanical properties, fracture toughness, and fracture risk. Full-spectrum and partial least squares (PLS) regression based models have been used to relate *in vivo*, ex vivo or exposed-bone Raman measurements to fracture toughness and other mechanical endpoints in human and murine bone [13], [26], [37]. More recently, Unal et al. [38] applied machine-learning regression models to ex vivo bone Raman spectra from 118 human femoral cortical specimens, achieving high fracture-toughness prediction accuracy ($R^2$ up to ~0.74). Similar strategies, together with broader machine-learning frameworks, have also been widely adopted across biomedical Raman applications, including disease classification and diagnosis [39], [40], [41], [42], [43]. In all these studies, however, machine learning is used to predict clinical or mechanical outcomes from spectra that are already dominated by bone signal, rather than to recover the bone spectrum itself from transcutaneous measurements.

In the present work, we treat bone signal extraction as a supervised learning problem by employing multi-response PLS regression to learn direct mapping from transcutaneous finger SORS spectra to corresponding reference spectra acquired from exposed bone. To our knowledge, this is the first demonstration of using a machine-learning model to directly reconstruct bone-like spectra from transcutaneous SORS data, providing a new route to robust bone-signal isolation for noninvasive bone-health assessment. Building on these capabilities, this study aims to bridge the translational gap between spectral feasibility and clinically established classifications of bone health. Using cadaveric hand specimens spanning the spectrum of bone health (normal, osteopenic, and osteoporotic), we acquired transcutaneous SORS finger measurements and systematically paired them with site-matched underlying phalanx exposed bone spectra to establish a comprehensive reference library. By integrating these data with supervised machine learning models, we extracted bone-specific biochemical signatures from transcutaneous measurements, classified subjects according to WHO bone-health categories, and predicted distal-radius DXA T-scores from peripheral (finger) Raman measurements. In addition, we performed preliminary *in vivo* measurements in volunteers to assess the consistency between cadaveric and living-tissue transcutaneous spectra, supporting the translational relevance of the cadaver-based model. Together, these results validate the feasibility of Raman-based categorization of bone health and establish a framework for extending the method to *in vivo* human studies. Ultimately, our findings highlight the promise of a compact, non-ionizing Raman platform as a clinically relevant complement to DXA, with potential to enable earlier detection and improved management of osteoporosis.

## 2 Results

### 2.1 Stable Measurement Region

Across all Raman-derived biochemical metrics, the central zone spanning approximately ±5–6 mm of the phalangeal midpoint consistently provides the strongest discrimination between normal, osteopenic, and osteoporotic bone. Fig. 1 summarizes the spatial sampling strategy along the phalanges (Fig. 1A), which phalanges used in this study (Fig. 1B) and illustrates the position-dependent (spatial) variation in exposed-bone Raman spectra and derived biochemical metrics. As shown in this figure, measurements acquired within approximately ±5 mm of the phalangeal midpoint (MM00) exhibit reduced spectral variability and enhanced separation between WHO diagnostic categories compared with more proximal or distal locations. Based on this observation, the MP05–MD05 window was selected as the reference region for all subsequent analyses.

The spatial dependence of Raman spectral features underlying this selection is illustrated using D2P1 as a representative example. Fig. 1C–E show category-averaged exposed-bone Raman spectra acquired at a 3-mm offset along the proximal–distal axis for normal (n = 5), osteopenic (n = 5), and osteoporotic (n = 3) specimens. While spectral variation is observed across positions, spectra within the MP05–MD05 region display consistent peak shapes and relative intensities across cadavers. This stability is reflected quantitatively in Fig. 1F–H, which summarize key Raman-derived metrics as a function of position, including mineral quality ($PO_4^{3-}/CO_3^{2-}$), mineral-to-matrix ratio ($PO_4^{3-}$/Amide III), and organic matrix heterogeneity, $CH_2$ full width at half maximum (FWHM). Values are reported as category-level means ± standard error of the mean (SEM), where SEM is calculated as the standard deviation divided by $\sqrt{N}$ and reflects the consistency of each spectral feature across cadavers within a category. While SEM provides an indication of inter cadaver variability, the primary rationale for selecting the MP05–MD05 region lies in its superior ability to distinguish Normal, Osteopenia, and Osteoporosis categories.

From a practical perspective, the width of this central window is also advantageous for translation to transcutaneous measurements. A ±5 mm sampling region around MM00 is sufficiently broad to tolerate uncertainties in anatomical landmark identification, reducing sensitivity to small placement errors during *in vivo* acquisition. Taken together, these findings establish MP05–MD05 as the most robust and translationally relevant region for representative bone spectroscopy and bone-health classification.

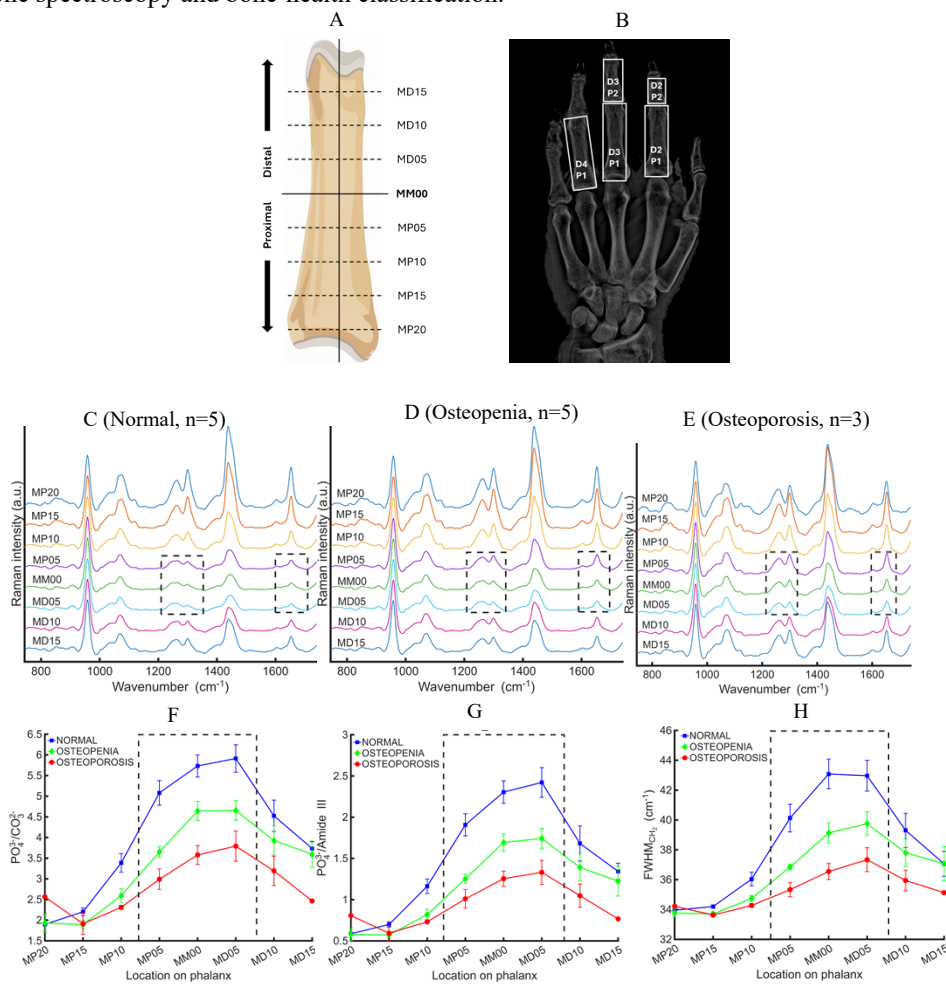

Fig. 1. Spatially resolved exposed-bone Raman measurements along the phalanges. (A) Schematic of measurement locations defined relative to the phalangeal midpoint (MM00). (B) X-ray showing the digits and phalanges analyzed. (C–E) Averaged exposed-bone Raman spectra measured along the proximal–distal axis of the D2P1 phalanx for (C) Normal, (D) Osteopenia, and (E) Osteoporosis categories; In all panels, *n* denotes the number of cadavers contributing to each group-averaged spectrum at a given spatial location (one spectrum per cadaver per position). Each trace represents the mean Raman spectrum at a specific position relative to MM00. (F–H) Mean ± SEM of key Raman metrics versus position: (F) $PO_4^{3-}/CO_3^{2-}$, (G) $PO_4^{3-}$/Amide III, and (H) $CH_2$ FWHM. Across all metrics, the MP05–MD05 region shows minimal variability and the strongest separation between diagnostic categories, supporting its use as the reference region for subsequent analyses.

## 2.2 ML-based Extraction of Finger Bone Raman Spectra

Following identification of MP05–MD05 as the reference region, all subsequent analyses were restricted to this window. Fig. 2 summarizes the performance of the machine-learning–based bone spectrum extraction within this region. Panels A–C compare category-averaged ML-predicted bone spectra (dashed black) with the corresponding measured 3-mm offset exposed-bone spectra (solid) for (A) Normal, (B) Osteopenia, and (C) Osteoporosis specimens. Across all diagnostic groups, the predicted spectra closely match the measured exposed-bone spectra, with high spectral overlap and strong correlation (correlation coefficients of 0.988 for Normal, 0.996 for Osteopenia, and 0.997 for Osteoporosis), demonstrates the model's ability to recover the bone-specific Raman features from transcutaneous measurements.

For reference, Fig. 2D–E show category-averaged spectra acquired from exposed bone at 0- and 3-mm collection offsets, respectively. These data illustrate that the 3-mm offset provides improved separation between WHO diagnostic groups compared with surface (0-mm) measurements. Fig. 2F presents the category-averaged ML-predicted bone spectra reconstructed from transcutaneous measurements. After normalization to the $PO_4^{3-}$ band peak (~960 cm$^{-1}$), the predicted spectra exhibit peak shapes, relative intensities, and category-dependent trends that closely resemble those observed in the 3-mm exposed-bone measurements, confirming that the extracted spectra retain diagnostically relevant bone-specific information. Including the exposed bone reference therefore provides a ground truth for assessing how well the SORS/ML-based prediction approach can discriminate WHO categories in practice.

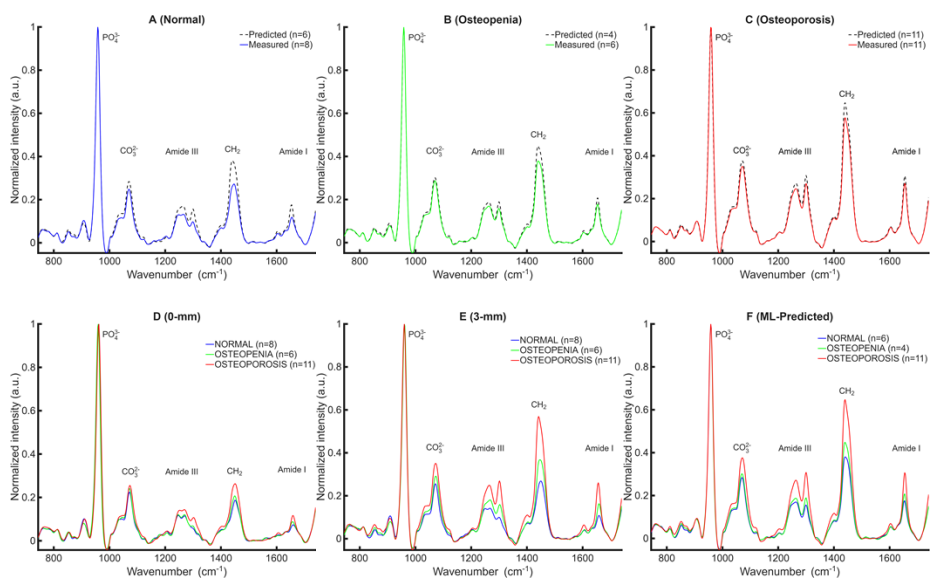

Fig. 2. (A–C) Comparison of category-averaged ML-predicted bone Raman spectra (dashed black), trained to predict exposed-bone spectra, with the corresponding measured 3-mm offset exposed-bone spectra (solid) from the midshaft (MM00) of D2P1, grouped by WHO diagnostic category: (A) Normal, (B) Osteopenia, and (C) Osteoporosis. In all panels, *n* denotes both the number of cadavers and the number of spectra contributing to each group average (one spectrum per cadaver). All spectra were normalized to the $PO_4^{3-}$ peak ($\approx$ 960 cm$^{-1}$) solely for spectral visualization and qualitative comparison. (D–E) Category-averaged exposed-bone spectra acquired at 0-mm and 3-mm collection offsets, illustrating enhanced separation of WHO groups with the 3-mm SORS geometry. (F) ML-predicted bone spectra reconstructed from 6-mm transcutaneous measurements, showing compositional differences that closely align with those observed in exposed bone.

## 2.3 Raman-Derived Spectral Metrics Analysis

### 2.3.1 Biochemical discrimination using Raman-derived metrics

Key Raman-derived biochemical metrics including mineral quality ($PO_4^{3-}/CO_3^{2-}$), and three mineral-to-matrix ratios ($PO_4^{3-}$/Amide III, $PO_4^{3-}/CH_2$, and $PO_4^{3-}$/Amide I) were evaluated across two data sources and shown in Fig. 3, exposed bone measured at 3-mm offset (Fig. 3A), and ML-predicted bone spectra extracted from transcutaneous measurements (Fig. 3B).

The 3-mm offset, corresponding to a deeper SORS geometry on exposed bone, enabling all metrics to discriminate normal from osteoporotic bone ($p \leq 0.001$), and also successfully separate intermediate groups. Analysis of ML-predicted bone spectra reveals that these overall biochemical trends are largely preserved. The predicted spectra reproduce the expected ordering of WHO categories, with higher mineral-to-matrix in normal bone relative to osteoporotic bone. However, statistical significance is retained only for the normal–osteoporosis comparison, while neither the normal–osteopenia nor osteopenia–osteoporosis comparisons achieve significance ($p \geq 0.05$). This indicates that the ML-based extraction captures the dominant compositional differences between normal and osteoporotic bone, whereas subtler variations between adjacent diagnostic categories are partially masked. This limitation is consistent with the exposed-bone results, where discrimination between adjacent categories emerges mostly at deeper (3-mm) offsets [32], suggesting that, in this transcutaneous configuration, the measurements provide limited access to the subsurface bone biochemical information needed to resolve these intermediate differences. The close agreement between the significant normal–osteoporosis contrasts in the predicted data and those obtained at the 3-mm offset from exposed bone indicates that the ML-based extraction model effectively recovers diagnostic features associated with bone mineral and matrix composition.

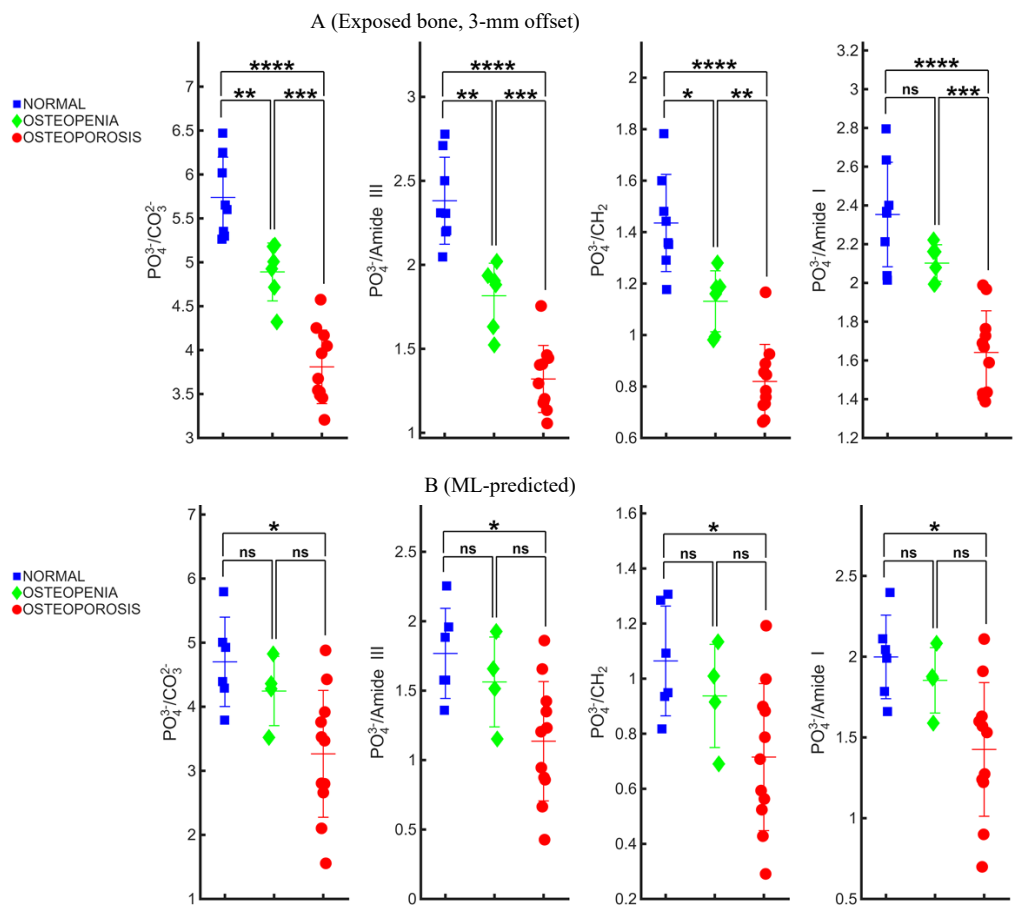

Fig. 3. Comparison of Raman-derived bone metrics across exposed and ML-predicted bone spectra. (A) Metrics derived from exposed bone measured at a 3-mm lateral offset, emulating a deeper SORS geometry on exposed bone. All metrics show enhanced discrimination between WHO categories, with normal–osteoporosis differences significant ($p \leq 0.001$) and consistent separation of adjacent groups, reflecting increased sensitivity to subsurface mineral–matrix composition. (B) Metrics computed from ML-predicted bone spectra extracted from transcutaneous measurements. The predicted spectra preserve the expected ordering across categories (Normal > Osteopenia > Osteoporosis) and retain significant discrimination for the normal–osteoporosis comparison across all metrics ($p \leq 0.05$), while differences between adjacent categories are not significant ($p \geq 0.05$). Overall, the ML-predicted spectra capture the principal biochemical trends observed in exposed bone, indicating that the extraction model preserves key subsurface diagnostic information accessible through SORS.

### 2.3.2 Prediction of distal-radius DXA T-scores from phalangeal Raman spectra

To evaluate whether phalangeal Raman spectra can serve as a surrogate for wrist bone mineral health, PLSR models were used to estimate distal-radius DXA T-scores from both exposed-bone spectra and ML–predicted bone spectra. Separate models were trained using spectra collected from midshaft (MM00) of the D2P1 at 3-mm offset from exposed bone, as well as the ML–predicted spectra extracted from transcutaneous measurements at the same location. Fig. 4A and B show the correlation between measured and predicted T-scores for these models. For the 3-mm offset exposed-bone spectra (Fig. 4A), the model achieves strong correlation of $r = 0.906$, $RMSE_{CV} = 0.832$ and the ML–predicted bone spectra derived from transcutaneous measurements (Fig. 4B) achieves performance of $r = 0.73$, $RMSE_{CV} = 1.38$, that are smaller than the 1.5-unit interval separating the primary WHO diagnostic thresholds (specifically, the span from the −1.0 cutoff for normal/osteopenia to the −2.5 cutoff for osteopenia/osteoporosis).

Categorical performance of the ML-predicted bone spectra is summarized in Fig. 4C, and D. The confusion matrix (Fig. 4C), derived from predicted T-scores using WHO thresholds, demonstrates accurate identification of osteoporotic and normal cases with limited misclassification between adjacent categories. Receiver-operating-characteristic (ROC) analysis (Fig. 4D) demonstrates discrimination against normal versus osteopenia (AUC = 0.75), normal versus osteoporosis (AUC = 0.97), and osteopenia versus osteoporosis (AUC = 0.77). Corresponding precision, sensitivity, specificity, and accuracy metrics are reported in Table 1. Collectively, these findings indicate that transcutaneous Raman measurements, when processed using the ML-based extraction model, retain quantitative predictive power comparable to that of exposed bone spectra, supporting the feasibility of noninvasive Raman spectroscopy for pre-screening or risk stratification prior to confirmatory DXA assessment.

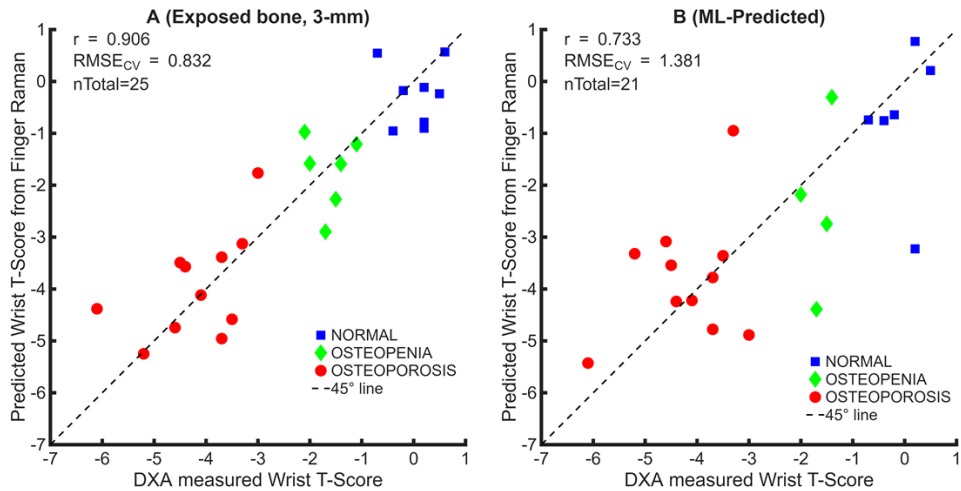

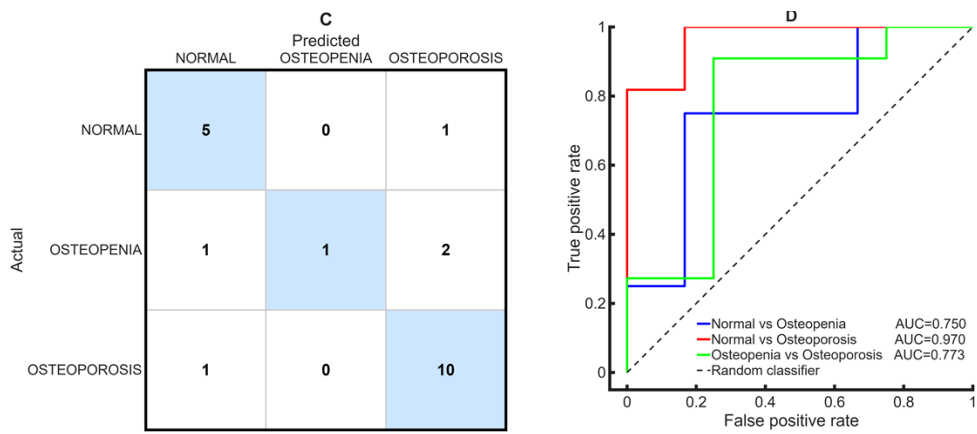

Fig. 4. Prediction of distal-radius DXA T-scores from phalangeal Raman spectra using PLSR with leave-one-subject-out cross-validation. (A, B) Measured versus predicted DXA T-scores derived from (A) 3-mm spatial-offset exposed-bone spectra, and (B) ML-predicted bone spectra extracted from transcutaneous measurements. Dashed lines indicate the line of identity; colors denote WHO diagnostic categories. (C) Confusion matrix for WHO category classification based on predicted T-scores. (D) Receiver-operating-characteristic (ROC) curves for normal versus osteopenia (AUC = 0.75), normal versus osteoporosis (AUC = 0.97), and osteopenia versus osteoporosis (AUC = 0.77) classification.

Table 1. Summary of classification performance metrics.

| WHO Class. | Precision | Sensitivity | Specificity | Accuracy |
| --- | --- | --- | --- | --- |
| Normal | 0.71 | 0.83 | 0.87 | 0.86 |
| Osteopenia | 1 | 0.25 | 1 | 0.86 |
| Osteoporosis | 0.77 | 0.91 | 0.7 | 0.81 |

## 2.4  Similarity to In Vivo Spectra

While cadaver-based data collection provides a controlled experimental framework for developing and validating machine-learning models, this does not ensure useful translation to *in vivo* applications. Tissue temperature, oxygenation saturation, and blood pulsation are among the most obvious ways that *in vivo* conditions differ from *ex vivo*. For spectral comparison, we therefore obtained preliminary *in vivo* Raman measurements by enrolling volunteer subjects under an IRB-approved protocol at the University of Rochester. Fig. 5 compares representative transcutaneous Raman spectra acquired at 6-mm spatial offset from two randomly selected cadaver fingers and two *in vivo* human fingers from volunteers (one male, 33 years old; one female, 28 years old). Spectra were normalized to the $CH_2$ band intensity to facilitate direct comparison of other relative peak features, particularly the phosphate band. Despite the above-mentioned differences in measurement conditions, the *in vivo* spectra replicated the spectral peak shapes and main ratios among all the major mineral and matrix peaks of the cadaveric transcutaneous measurements.

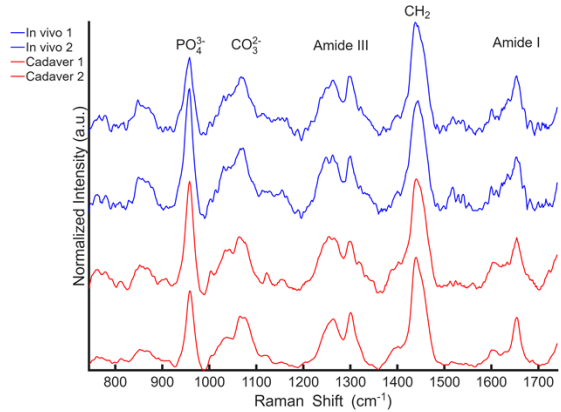

Fig. 5 Representative transcutaneous Raman spectra acquired at a 6-mm spatial offset from two cadaver fingers (Cadaver 1, and 2) and two volunteers (*In vivo* 1, and 2). All spectra are normalized to the $CH_2$ band intensity to enable direct comparison of relative peak features. Prominent bone- and matrix-associated bands are observed across both datasets, including phosphate ($PO_4^{3-}$), carbonate ($CO_3^{2-}$), Amide III, $CH_2$, and Amide I, demonstrating qualitative agreement between cadaveric and *in vivo* transcutaneous measurements and supporting the feasibility of SORS-based Raman assessment in living subjects.

## 3 Discussion

Current clinical management of osteoporosis relies heavily on DXA bone scans [5], yet these bone scans fail to explain fracture risk because they measure mineral density while overlooking critical chemical and molecular determinants of bone quality and fragility fracture risk [44]. While Raman spectroscopy offers a molecular fingerprinting alternative for assessing bone quality metrics, such as mineral-to-matrix ratios and mineral quality, its clinical utility has been limited by the technical challenge of decomposing weak bone signals from the spectral interference of overlying skin and soft tissue. In this cadaveric study, we addressed this translational gap by validating a supervised machine learning (Multi-response PLS Regression) framework to predict intrinsic bone spectra directly from transcutaneous SORS measurements of the finger. Using paired transcutaneous and exposed bone SORS data from 25 female cadaveric hands, our supervised ML model successfully reconstructed bone-specific signatures. Importantly, the ML-predicted finger spectra retained quantitative diagnostic power, enabling estimation of distal radius T-scores and statistically significant discrimination between normal and osteoporotic subjects.

Progress in transcutaneous Raman spectroscopy of bone has historically followed two parallel tracks: (i) instrument/probe innovations that increase subsurface sensitivity, and (ii) computational unmixing of transcutaneous measurements to recover bone-specific spectra. On the instrumentation side, early human demonstrations employed SORS to obtain the first *in vivo* bone spectrum from the distal thumb phalanx under safe irradiances, establishing feasibility but providing limited quantitative validation due to the lack of a subject-matched ground truth [25]. On the computational side, most prior efforts have relied on spectral unmixing techniques such as BTEM, adaptive BTEM, PARAFAC, and SOLD to isolate bone features from transcutaneous spectra. These approaches decompose transcutaneous spectra into bone and soft-tissue components using constrained factor rotations and entropy minimization (BTEM-type methods), multi-way factorization (PARAFAC), or over constrained library fitting (SOLD) [33], [34], [35], [36]. While appealing because they require minimal or no paired data, the recovered bone component is not explicitly tied to a site- and subject-matched exposed bone reference spectrum, and performance therefore depends on the suitability of soft-tissue libraries and assumptions regarding tissue composition.

Although existing studies have established proof-of-concept for noninvasive bone spectroscopy, to our knowledge, no prior studies have successfully extended these approaches to clinical classification according to the bone health categories of normal, osteopenia, and osteoporosis, as defined by DXA T-score thresholds. Even in the few human studies that have attempted transcutaneous group-level comparisons, such as the work of Buckley et al. [26], the extracted bone spectra did not yield statistically significant differences between osteoporotic and control cohorts despite exhibiting the expected trend in mineral-to-matrix composition, highlighting the difficulty of achieving reliable diagnostic discrimination through skin, given that stratification into these diagnostic categories directly informs fracture-risk management and therapeutic decision-making.

Our current approach directly addresses this limitation by constructing a paired transcutaneous and exposed bone spectral library from identical anatomical sites and training a supervised machine-learning model to learn an explicit mapping from the mixed transcutaneous signal to the corresponding exposed-bone spectrum. Unlike prior unmixing strategies, this framework does not require a soft-tissue reference library and instead uses site- and subject-matched exposed bone as ground truth. By incorporating specimens spanning all WHO diagnostic categories, the model enables direct reconstruction of bone-specific spectral signatures that retain quantitative discrimination between bone-health groups. Importantly, to our knowledge, this work represents the first demonstration of using supervised machine learning to (a) directly reconstruct bone-like Raman spectra from transcutaneous SORS measurements and (b) leverage those reconstructed spectra for statistically significant classification of bone health according to WHO categories, rather than predicting downstream mechanical or clinical endpoints from directly measured bone spectrum. While our model successfully differentiated normal from osteoporotic bone, the spectral discrimination of the intermediate osteopenic group remains challenging. This suggests that while the dominant spectral changes associated with severe disease are recoverable transcutaneously, subtler mineral-matrix gradations may require larger training cohorts or further improvements in experimental setup or reconstruction algorithms to achieve statistical significance. In addition, a practical outcome of this study is the identification of an approximately 1-cm mid-phalangeal sampling window (MP05–MD05) in which biochemical variability is minimized and diagnostic separation is maximized. This spatial robustness substantially relaxes alignment requirements for translation to clinical and point-of-care applications.

To further benchmark our approach, Table 2 compares our bone-signal extraction accuracy with that of established decomposition methods. Our model achieves a Pearson correlation of 0.995 between predicted and ground-truth bone spectra from 25 human cadaver hands with a wide range of T-scores from healthy normal to severely osteoporotic, matching or approaching the performance of A-BTEM (>0.996 on one human cadaver leg), SOLD (0.996 in mice tibiae), and classical BTEM (0.96 in mice tibiae). The key distinction lies in methodology: BTEM, A-BTEM, and SOLD require a multi-component spectral library consisting of reference bone and soft-tissue spectra and explicitly unmix the transcutaneous measurement through entropy minimization or constrained factorization. Our approach eliminates the need for soft-tissue reference spectra, which vary substantially across anatomical sites and subjects, and instead learns a direct regression from transcutaneous to bone spectrum using only exposed bone as ground truth. This reduces sensitivity to soft-tissue heterogeneity and supports more robust generalization across donors.

Table 2. Comparisons of the bone signal extraction method from this work with other methods previously reported. Notably, the proposed machine-learning approach achieves high correlation (r = 0.995) without needing a separate soft-tissue reference library, unlike the other listed decomposition techniques.

| Bone spectra extraction method | Pearson's correlation coefficient | Comment | Ref. |
|---|---|---|---|
| Machine learning | 0.995 | 21 human cadaver hands, 830nm, 100 mW | This work |
| A-BTEM | > 0.996 | 1 human cadaver leg, 830nm, 320 mW | [36] |
| SOLD | 0.996 | 21 mice tibiae , 830nm, 150 mW | [34] |
| BTEM | 0.96 | 32 mice tibiae, 785nm, 400 mW | [33] |

Several large clinical studies and consensus reports have established that standard imaging-based techniques provide clinically useful but inherently limited discrimination for osteoporosis and fracture risk. DXA-derived areal BMD, while the clinical reference standard, is known to exhibit substantial overlap between fracture and non-fracture populations and reduced sensitivity for identifying individuals who will go on to experience fragility fractures, particularly within osteopenic cohorts [45], [46]. Quantitative computed tomography (QCT) has been reported to enhance osteoporosis detection in some populations by providing volumetric density information; however, this comes with increased radiation exposure, reduced accessibility, and incomplete diagnostic concordance with DXA [47], [48]. In this context, the present Raman–ML framework demonstrates classification performance for WHO categories while simultaneously achieving statistically significant separation between normal and osteoporotic groups across multiple Raman-derived biochemical parameters, including several mineral-to-matrix ratios. Clinically, this approach provides a non-ionizing radiation complement to DXA, capable of capturing bone compositional markers at the point of care. The correlation between Raman-derived metrics predicted from finger measurements and DXA T-scores at clinically relevant fracture sites, such as the wrist, together with discrimination between healthy and osteoporotic subjects, supports the potential utility of this method for screening individuals at risk of fragility fractures. Moreover, the portability and favorable safety profile of Raman spectroscopy position this technology for longitudinal monitoring, community-based screening, and eventual deployment in primary-care settings.

This study has several limitations that warrant consideration. First, while the cadaveric model provides an indispensable subject-matched ground-truth framework for validating bone-signal extraction, it does not fully replicate the physiological conditions of living tissue, including pulsatile blood flow, temperature gradients, and dynamic hydration, which may introduce additional spectral variability in clinical measurements. Second, laser safety constraints for *in vivo* skin exposure required reducing the excitation power from 100 mW used in cadaver experiments to approximately 30 mW in human volunteers. Although comparable photon dose and signal-to-noise ratio were achieved by summing multiple acquisitions, this highlights the need for further optimization of collection efficiency and acquisition time to support practical clinical throughput. Third, the cadaveric cohort was modest in size (N = 21) and predominantly White, and the preliminary *in vivo* validation was limited to a small number of volunteers. Future studies will be required to validate this approach in larger and more diverse populations, including osteoporotic patients and individuals with a broad range of skin pigmentation, as melanin content may influence fluorescence background and effective optical penetration depth in transcutaneous Raman measurements. Finally, while multi-response PLS regression proved robust for the present dataset, larger datasets may enable exploration of nonlinear or deep learning architectures to further improve bone-spectrum reconstruction and diagnostic performance.

## 4 Materials and methods

### 4.1 Cadaver samples and bone preparation

Cadaveric hands from 25 female donors, obtained in two cohorts (12 in the first and 13 in the second), were provided by the Anatomy Gifts Registry (Hanover, MD, USA). Inclusion criteria for donors included HIV-negative serology, less than 3-year postmortem recovery, and available medical history. Exclusions include known arm fractures, nonambulatory for more than 1-year, musculoskeletal impairments (paralysis or paresis), neoplasm, or prosthetic hardware. Donor characteristics age, body mass index (BMI), ethnicity, 1/3 radius BMD, and T-score are summarized in Table 3. The cohort had a mean ± SD age of 70 ± 15 years (range 40–99 years) and was predominantly White (24 White, 1 Black). DXA scans of the distal 1/3 radius were acquired by Advanced Radiology (Hanover, MD, USA) using a Horizon Ci DXA system (Hologic). Based on T-scores, the exposed-bone analysis included n = 8 donors classified as Normal, n = 6 as Osteopenic, and n = 11 as Osteoporotic. Four specimens could not be used for transcutaneous SORS analysis due to measurement issues that could not be repeated post-dissection. After excluding these samples, the transcutaneous dataset comprised n = 6 Normal, n = 4 Osteopenic, and n = 11 Osteoporotic donors (total n = 21).

All cadavers were stored at −80 °C on arrival. Prior to Raman measurements, cadaver hands were thawed for at least 12 hours before transcutaneous scanning. Following transcutaneous measurements, the required phalanges were carefully dissected, and after removing the surrounding soft tissues, the exposed bone specimens were rehydrated for 2 hours in 1× phosphate-buffered saline (PBS) at room temperature. Specimens were then wrapped in 1x PBS–soaked gauze to maintain hydration until Raman acquisition occurred. Upon completion of Raman acquisition, specimens were returned to the −80 °C freezer for future analyses.

Table 3. Demographics of distal-radius cadaver donors (mean ± SD).

| WHO Class. | Ethnicity | Age | BMI | 1/3 Radius BMD | 1/3 Radius T-score |
|---|---|---|---|---|---|
| | (Black/White) | (years) | (kg/m$^2$) | (g/cm$^2$) | |
| Normal | 1 / 7 | 56.5 ± 10.1 | 31.4 ± 6.4 | 0.698 ± 0.028 | 0.05 ± 0.45 |
| Osteopenia | 0 / 6 | 69.0 ± 7.6 | 24.7 ± 5.3 | 0.556 ± 0.023 | −1.63 ± 0.38 |
| Osteoporosis | 0 / 11 | 82.5 ± 9.9 | 21.2 ± 4.2 | 0.443 ± 0.054 | −4.19 ± 0.90 |

### *4.2 SORS setup*

The SORS measurement strategy and probe architecture build upon prior systems developed in our laboratory for Raman-based evaluation of bone in the human hand [15], [31], [32], while the optical layout and collection geometry were specifically adapted for the objectives of the present study. Three spatially offset detection channels were implemented at lateral distances of 0, 3, and 6 mm from the laser excitation point. This geometry enables depth-selective Raman sampling, with increasing offset corresponding to greater photon migration depth within the tissue, as illustrated schematically in Fig. 6A. Larger offsets preferentially enhance subsurface bone contributions while suppressing superficial soft-tissue signals. At each offset distance (0, 3, and 6 mm), five consecutive Raman frames were acquired with 60-s integration time per frame and averaged to yield a single high-signal-to-noise spectrum.

The optical configuration used for both cadaveric and in vivo measurements is shown schematically in Fig. 6B. Raman excitation was delivered to the sample surface, and scattered light was collected through spatially separated fiber bundles conjugated to the defined lateral offsets.

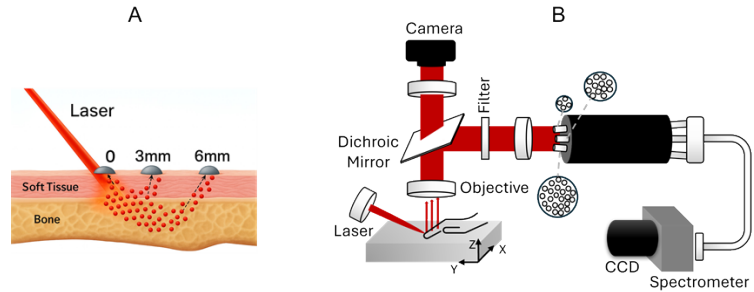

Fig. 6. Spatially Offset Raman Spectroscopy (SORS) geometry and instrumentation. (A) Conceptual illustration of spatially offset Raman geometry showing increasing photon penetration depth at larger lateral offsets (0, 3, and 6 mm). Larger offsets preferentially enhance subsurface bone signal relative to overlying soft tissue. (B) Schematic of the optical configuration used for SORS measurements, including excitation delivery and spatially separated fiber-bundle detection channels. Three fiber bundles, containing 4, 12, and 26 fibers respectively, are physically spaced apart on the probe end so that, once aligned using back-illumination from the spectrograph, they collect Raman signals at 0-, 3-, and 6-mm offsets from the laser spot on the sample.

The SORS instrumentation and optical alignment procedures used in this study have been described in detail in our recent publication [32]. Briefly, a continuous-wave 830-nm diode laser (Process Instruments Inc.) (100 mW for cadaveric study and 30 mW for *in vivo* study) was directed onto the sample surface through a 50-mm focal-length objective (NA ≈ 0.29) at an incidence angle of approximately 60° relative to the surface normal, producing an elliptical excitation spot (~1 mm minor axis). Raman-scattered light was collected normal to the surface using a 30-mm focal-length objective (NA ≈ 0.39) and directed through a dichroic mirror into three fused-silica fiber bundles positioned to correspond to 0-, 3-, and 6-mm spatial offsets at the sample surface. The bundles were configured to balance signal intensity across offsets, containing 4, 12, and 26 fibers respectively.

The collected light was delivered to a Kaiser Optical Systems HoloSpec VPT imaging spectrograph (f/1.8) equipped with a thermoelectrically cooled Andor iDus DU420A-BEX2-DD deep-depletion CCD (1024 × 256 pixels, 26 μm pitch, operated at −55 °C). Specimens were mounted on a motorized XYZ translation stage (10-μm Z-resolution) to ensure consistent anatomical positioning. Optical alignment, offset calibration, and probe geometry were verified using back-illumination and real-time imaging of the laser spot.

## 4.3 Data collection

### 4.3.1 Cadaver measurements

For each phalanx, the total dorsal length from proximal joint to distal joint was measured, and the midpoint was marked as MM00. All other spatial locations were defined relative to this reference (e.g., MP20 = 20 mm proximal to MM00; MD15 = 15 mm distal to MM00). SORS measurements were acquired at predefined intervals along the dorsal surface of each phalanx, referenced to its anatomical midpoint (denoted MM00). Cadavers were received in two cohorts ($n = 12$ and $n = 13$). For the first cohort, measurements were collected at 3-mm increments (e.g., MP03, MP06), whereas for the second cohort, 5-mm increments were used (e.g., MP05, MP10). Data acquisition proceeded sequentially from the most proximal location toward the midpoint and then continued distally toward the distal end of the phalanx. At each spatial location, the laser excitation spot and 0-mm collection fiber were repositioned and centered on the designated coordinate, such that the measurement geometry moved along the phalanx with each step. This proximal-to-distal sequence ensured consistent spatial registration between transcutaneous and subsequent exposed-bone measurements. Measurements were acquired on both transcutaneous fingers and, following dissection, on the corresponding exposed bones at matched anatomical sites. To increase dataset size for supervised machine-learning analysis, measurements were collected across multiple digits, including D2P1, D3P1, D4P1, D3P2, and D2P2 (Fig. 1B).

For all cadaveric measurements (both transcutaneous and exposed bone), the laser power was 100 mW. At each spatial location and for each offset (0, 3, and 6 mm), five consecutive Raman frames were acquired with 60-s integration time per frame and averaged to generate a single high signal-to-noise spectrum.

For cross-cadaver comparative analyses, ratio calculations, and distal-radius T-score prediction, measurements from the midpoint of the proximal phalanx of the second digit (D2P1-MM00) were used as the representative site for each cadaver. Measurements from additional phalanges and neighboring positions within the MP05–MD05 reference window were incorporated exclusively to expand the training dataset for the machine-learning model.

### 4.3.2 In vivo measurements

*In vivo* transcutaneous Raman spectra were acquired at the midpoint of the proximal phalanx of D2P1 using a SORS geometry consistent with the cadaver studies, with laser power limited to approximately 30 mW to comply with laser safety requirements for skin exposure as defined by BS/IEC 60825-1. Because cadaver measurements were performed at a higher excitation power (100 mW), a fair comparison between cadaveric and *in vivo* spectra was achieved by summing three consecutive *in vivo* acquisitions, thereby matching the total delivered photon dose to that of a single cadaver measurement.

## 4.4 Data processing and analysis

### 4.4.1 Data processing

All spectral preprocessing and calibration were performed in MATLAB R2025b using an in-house analysis pipeline and GUI. Detailed descriptions of wavelength calibration, detector-response correction, fiber normalization, cosmic-ray removal, and fluorescence background subtraction have been reported previously in our recent exposed bone study [32]. The fully processed spectra were then used for subsequent quantitative and machine-learning analyses.

### 4.4.2 Machine learning

In this supervised machine learning approach, Raman spectra obtained from transcutaneous measurements were paired with their corresponding exposed bone spectra from the same cadaver and location, but only within the MP05–MD05 region. This pairing enables the model to learn a mapping that isolates the underlying bone signal from overlying soft tissue contributions. The model was trained to predict 3-mm offset exposed-bone spectra from the 6-mm transcutaneous measurements using supervised Multi response PLS (PLS2) regression.

For the supervised machine learning section, for each spectrum we constructed a predictor vector $x_i \in \mathbb{R}^P$ from the transcutaneous SORS measurements (only 6-mm offset) and a response vector $y_i \in \mathbb{R}^Q$ from the corresponding exposed bone spectrum at the same spatial location (both with $Q = P = 499$ wavenumber points). Stacking all spectra from the training subjects yielded predictor and response matrices $X_{train} \in \mathbb{R}^{N_{train} \times P}$ and $Y_{train} \in \mathbb{R}^{N_{train} \times Q}$. We then used multi-response partial least squares regression (PLS2) to learn a mapping from the transcutaneous spectra to their corresponding exposed bone spectra [49], [50]. PLS2 decomposes these matrices into a low-dimensional latent space [51]: $X_{train} = TP^\top + E$, $Y_{train} = UQ^\top + F$, Here, $T, U \in \mathbb{R}^{N_{train} \times A}$ are the score matrices; $P, Q$ are the loading matrices; and $E, F$ are the residuals. The score vectors $t_a$ and $u_a$ are chosen iteratively to maximize their covariance under orthogonality constraints. The resulting regression model relates transcutaneous and bone spectra via $Y \approx 1 b_0^\top + XB$, with coefficient matrix $B \in \mathbb{R}^{P \times Q}$ and intercept $b_0 \in \mathbb{R}^Q$. In practice we used MATLAB's *plsregress* function [52], which returns an augmented coefficient matrix $\tilde{B} \in \mathbb{R}^{(P+1) \times Q}$ such that predictions for a test set $X_{test}$ are obtained as $\tilde{Y}_{predicted} = [1_{N_{test}} \quad X_{test}] \tilde{B}$.

Evaluation followed a leave-one-subject-out (LOSO) protocol. In each iteration, all spectra from one cadaver were held out as the test set, while the spectra from the remaining cadavers comprised the training set. This subject-wise cross-validation approach prevents data leakage arising from repeated measurements within the same specimen and is recommended for biomedical spectroscopy and multivariate modeling where samples from the same subject are not statistically independent [53], [54]. A critical hyperparameter in PLS regression is the number of latent components ($A$). In a preliminary leave-one-out iteration analysis, we performed grouped $K$-fold cross-validation (with folds defined by subject identity). This analysis revealed that the optimal model complexity was highly consistent across subjects, with the cross-validation error minimizing at approximately 10 components (range: 9–11). To ensure model stability and avoid overfitting specific training folds, we fixed the number of latent components to $A = 10$ for all subsequent leave-one-out iterations.

### 4.4.3 T-score prediction modeling

For cross-cadaver visualization and distal-radius T-score prediction, we used the anatomically consistent site D2P1-MM00, which was available for all cadavers. For T-score prediction modelling, a LOSO PLSR was implemented with the *plsregress* function in MATLAB. Processed spectra served as predictors, while distal-radius T-scores were responses. Model rank (1–15) was chosen in each iteration, and performance was quantified by the Pearson correlation coefficient ($r$) and the root-mean-square error of cross-validation (RMSE$_{CV}$). For each model rank, LOSO was performed, and the optimal rank was selected as the one yielding the minimum RMSE$_{CV}$.

Model performance was quantified using commonly used classification metrics derived from the confusion matrix. For each WHO class, the entries of the confusion matrix were used to obtain true positives (TP), true negatives (TN), false positives (FP), and false negatives (FN). Accuracy represented the overall proportion of correctly classified samples, $(TP+TN)/(TP+FP+TN+FN)$, Precision quantified the fraction of predicted positives that were correct, $TP/(TP+FP)$, while sensitivity (recall) measured the proportion of true positives correctly identified, $TP/(TP+FN)$, Specificity quantified the true-negative rate, $TN/(TN+FP)$.

**Funding.** Grant numbers R01AR070613 (AB and HA), and P30AR069655 (HA) from NIAMS/NIH.

**Acknowledgments.** The study was supported by grant numbers R01AR070613 (AB and HA), and P30AR069655 (HA) from NIAMS/NIH. The content is solely the responsibility of the authors and does not necessarily represent the official view of the NIH.

**Disclosures.** The authors declare that they have no known competing financial interests or personal relationships that could have appeared to influence the work reported in this paper.

**Data Availability.** Data underlying the results presented in this paper are available from the corresponding authors upon reasonable request